\documentclass[aip,graphicx,reprint]{revtex4-1}

\usepackage[dvipdfmx]{graphicx}
\usepackage{dcolumn}
\usepackage{bm}

\usepackage[utf8]{inputenc}
\usepackage[T1]{fontenc}
\usepackage{mathptmx}
\usepackage{etoolbox}
\usepackage{float}
\usepackage{siunitx}
\usepackage[version=4]{mhchem}
\usepackage{comment}

\newcommand{\mk}[1]{\textcolor{black}{#1}}
\newcommand{\sn}[1]{\textcolor{black}{#1}}
\newcommand{\mkk}[1]{\textcolor{black}{#1}}

\draft 

\begin{document}


\title{Statistical study and parallelisation of multiplexed single-electron sources} 



\author{S. Norimoto} 
\affiliation{National Physical Laboratory, Hampton Road, Teddington TW11 0LW, United Kingdom}
\author{P. See} 
\affiliation{National Physical Laboratory, Hampton Road, Teddington TW11 0LW, United Kingdom}
\author{N. Schoinas} 
\affiliation{National Physical Laboratory, Hampton Road, Teddington TW11 0LW, United Kingdom}
\author{I. Rungger} 
\affiliation{National Physical Laboratory, Hampton Road, Teddington TW11 0LW, United Kingdom}
\author{T. O. Boykin II}
\affiliation{Physical Measurement Laboratory, National Institute of Standards and
Technology, Gaithersburg, MD 20899, United States of America}
\affiliation{\sn{Joint Quantum Institute, University of Maryland, College Park, MD 20742, United States of America}}
\author{M. D. Stewart Jr}
\affiliation{Physical Measurement Laboratory, National Institute of Standards and
Technology, Gaithersburg, MD 20899, United States of America}
\author{J. P. Griffiths} 
\affiliation{Cavendish Laboratory, University of Cambridge, J. J. Thomson Avenue, Cambridge CB3 0HE, United Kingdom}
\author{C. Chen} 
\affiliation{Cavendish Laboratory, University of Cambridge, J. J. Thomson Avenue, Cambridge CB3 0HE, United Kingdom}
\author{D. A. Ritchie} 
\affiliation{Cavendish Laboratory, University of Cambridge, J. J. Thomson Avenue, Cambridge CB3 0HE, United Kingdom}
\author{M. Kataoka}
\affiliation{National Physical Laboratory, Hampton Road, Teddington TW11 0LW, United Kingdom}



\date{\today}

\begin{abstract}
Increasing electric current from a single-electron source is a main \mk{challenge in an effort} to establish the standard of the ampere defined by \mk{the fixed value of the} elementary charge $e$ and operation frequency $f$.
\mk{While the current scales with $f$, due to} an operation frequency limit \mk{for maintaining accurate single-electron transfer}, parallelisation of single-electron sources is expected to be a \mk{more practical} solution to increase \mk{the generated} electric current $I=Nef$, where $N$ is a number of \mk{parallelised} devices.
\mk{One way to} parallelise single-electron sources \mk{without increasing the complexity in device operation is to use} a common gate.  \mk{Such a scheme will require} each device to have the same operation parameters \mk{for single-electron transfer}.
\mkk{In order to investigate this possibility}, we study the statistics for operation gate voltages using single-electron sources embedded in a multiplexer circuit.
The multiplexer circuit \mk{allows} us \mk{to} measure 64 single-electron sources individually in a single cooldown. 
We also \mk{demonstrate} the parallelisation of three single-electron sources and \mk{observe} the generated current \mk{enhanced} by a factor of three.

\end{abstract}

\maketitle 

The 2019 redefinition of the SI units has fixed the numerical \mk{value of the} elementary charge $e$ and \mk{the} Planck constant $h$, which has established the Josephson effect and quantum Hall effect as independent realisation methods for \mk{the SI} volt and ohm derived from these fundamental physical constants, respectively~\cite{bipm2019}.
The ampere, one of the seven base units in SI, still lacks a direct realisation method and \mk{has} to be derived from the volt and ohm via Ohm's law in practice.
While this arrangement does not create hindrances in a majority of applications in the present metrology system, it struggles to provide high precision in the limit of small electric current, at which niche applications exist~\cite{Brun2016}.
In this limit, a direct realisation method by single-electron transfer may be advantageous.

Single-electron sources have been studied for decades in an effort~\cite{jukka2013, Kaneko_2016} to establish a standard of the ampere $I$ defined by elementary charge $e$ and transfer rate $f$, $I=ef$.
\mk{One milestone for \mkk{a} practical single-electron standard would be to match its} uncertainty \mk{to the one} derived \mk{from quantum Hall resistance and Josephson voltage} standards.
\mk{However, the generated current by a state-of-the-art single-electron source
is at a level of 100~pA,} and is too small to \mk{be measured} with an uncertainty of \qty{10}{nA/A} \mk{even with the existing capabilities of metrology laboratories (for example, the room-temperature Johnson noise of the feedback resistor in the transimpedance amplifier would limit the measurement uncertainty that can be achieved.)}.
Ideally, \mk{for the usage} as a practical current standard, a single-electron source must generate a current \mk{at least} \qty{1}{nA} to complete a measurement within a \mk{reasonable timescale (e.g. less than one day)}.
One strategy \mk{would be to increase} the operation frequency \mk{to} around \qty{6}{GHz}. 
\mk{However, the operation frequency beyond 1~GHz tends to reduce the accuracy of single-electron transfer}.\cite{Kataoka_2011}
The highest accuracy and operation frequency reported for single-device operation has been \qty{20}{\mu A/A} at \qty{7}{GHz}~\cite{Yamahata2017-nw} or \qty{0.2}{\mu A/A} at \qty{2}{GHz}~\cite{Giblin_2023}.
Another strategy \mk{would be to parallelise} single-electron sources, a similar idea to the Josephson voltage standards as many Josephson junctions are serialised to realise a volt.
Parallelisation of $N$ devices \mkk{multiplies} the current by a factor of $N$, $I=Nef$.
Single-electron source parallelisation has been demonstrated using SINIS junctions~\cite{Maisi_2009,shuji2015} but \mk{this particular} operation scheme, turnstile operation, has a disadvantage \mk{that the accuracy of current quantisation is lost} at high operation frequency (above order of \qty{100}{MHz}).
Recently, an article reported that single-electron source parallel operation increased the generated current with operation frequency at \qty{1}{GHz} but each device is driven by an individual gate electrode~\cite{Nakamura2023-tx}.
\mk{In order to integrate a large number of devices for parallelisation, a use of common gates\cite{Mirovsky_2010,kim2022} are preferred in terms of simplifying the device operation and signal wiring}.
This multiplication requires a condition that each device shares an operation condition to generate the current of $1ef$.
It is \mk{therefore important to investigate the} statistics \mk{of} the operation conditions \mk{for current quantisation in a large number of devices}.
\mk{However, the characterisation of single-electron sources at a dilution refrigerator temperature is time consuming, and an} elegant approach is required to acquire data \mk{large} enough \mk{for valid statistics}. 

In order to \mk{enable the characterisation of a large number of} single-electron sources efficiently, we designed and fabricated a \mk{set of devices that have} 64 single-electron sources embedded in a multiplexer circuit~\cite{al2013}.
\mk{These devices allow} us to measure a large number of electron pumps in a single cooldown.
We \mk{investigated the statistics of the operation conditions for a single-electron source to generate} the current of $1ef$.
The multiplexer circuit allows us to access each device individually, \mk{as well as to parallelise} a certain combination of \mk{multiple} devices.
We found a group of single-electron sources that showed current quantisation under the same device operating parameters (by coincidence), and \mk{demonstrated the parallelisation of} three single-electron sources, \mk{enhancing} the current by a factor of three.

Figure~1(a) shows schematics of a multiplexer circuit (only three layers are shown as an example, but the actual device has six layers).
There are two choices, left or right, to select a branch on each layer by applying gate voltage on gate electrodes (L\textit{i} and R\textit{i}, where \textit{i} is an integer number to specify a layer).
For convenience, \mkk{we} use a binary number to represent the choice at each layer as 0 for left and 1 for right.
Multiplexer circuit with \textit{n} layers \mkk{provides access to $2^{n}$ devices}.
We use a six-digit binary number to specify the state of multiplexer ($d_{1}d_{2}...d_{n}$ where $d_{i}$ is a choice at \textit{i}th layer) as the state to access \mkk{the $N$th device,} expressed by a binary number of $N-1$.
For example, a three-layer multiplexer circuit can access 3rd device with multiplexer state $3-1 = 0d2 = 0b010$.
The binary notation has an advantage when you choose both branches at a layer by setting both channels open (we note the choice as X).
Multiplexer in the state "X00" provides access to 1st device and 5th device simultaneously.
\mkk{The} multiplexer circuit provides \mkk{the capability} to measure single-electron sources individually and to parallelise \mkk{certain combinations} of them.

Figure~1(b) shows an optical photo of single-electron sources embedded in a six-layer multiplexer circuit.
A single-electron source~\cite{Blumenthal2007-ki, Giblin2012-wf} is embedded at the end of each downstream.
The device was fabricated on a \ce{GaAs/AlGaAs} heterostructure substrate \mkk{containing a two-dimensional electron gas (2DEG)} using conventional optical and electron beam lithography techniques except for cross-linked poly methyl methacrylate (PMMA) layers made by electron beam overdose~\cite{al2013}.
Cross-linked PMMA layers underneath gate electrodes make their pinch-off gate voltages \qty{0.7}{V} stronger than normal pinch-off voltage around \qty{-0.35}{V}, which enables \mkk{the development of the} multiplexer circuit~\cite{al2013}.
The tree of our multiplexer circuit starts from the top right source Ohmic contacts (marked S) towards the drain contact (marked D) in the middle.
The circuit consists of six layers and has twelve gate electrodes, L1-6 and R1-6, to select active channels.
In total, $2^{6} = 64$ single-electron sources are \mk{arranged} in a square \mk{surrounding} the drain contact.
This multiplexer method obviously has an advantage in requiring a much smaller number of gates to \mk{control a large number of devices} compared to a method using individual gates~\cite{kim2022}.
All single-electron sources are fabricated by the same design and share Entrance and Exit gates.
\begin{figure}[htbp]
\centering
\includegraphics[width=3.2in]{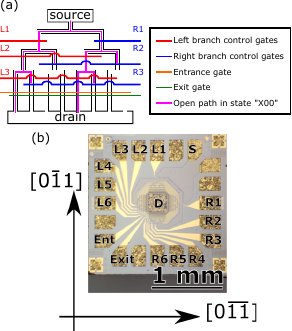}
\caption{(a) Schematic of three-layer multiplexer circuit. Gate electrodes (L$i$ and R$i$) on a multiplexer circuit provide two choices of left (0) and right (1) at each layer. Selecting both branches (X) enables to address multiple devices at the same time. For example, a multiplexer in state "X00" allows to address device "000" and device "100" at the same time (pink line).
(b) An optical photo of our device. The device has a six-layer multiplexer circuit so that it provides access to $2^{6} = 64$ devices in total. A single-electron source is embedded in each branch at the end of downstream and all single-electron sources share \mk{the} Entrance and Exit gates.\label{device}}
\end{figure}
Measurement was carried out using a dilution refrigerator with a base temperature of \qty{20}{mK}.
\mkk{A magnetic field up to 12.5~T was applied in the direction perpendicular to the plane of the 2DEG.}
Gate electrodes for the multiplexer (L1-6 and R1-6) were connected to two eight-channel voltage sources through homemade 1/10 voltage dividers and low pass filters.
Each branch is pinched off by applying a gate voltage of \qty{-0.4}{V} on the gate electrode.
Entrance and Exit gates to manipulate single electron sources are connected to \mk{dc} voltage sources \mk{outputting $V_{\textrm{ent}}$ and $V_{\textrm{exit}}$, respectively,} through homemade low pass filters and \mkk{bias-tees}.
Microwaves to operate single electron sources are generated by Tektronix AWG70001A synchronised with \qty{10}{MHz} reference signal.
The microwave reaches the Entrance gate through a low pass filter, a radio frequency (RF) amplifier, attenuators, \mk{and is added to $V_{\textrm{ent}}$ via a bias-tee at the low temperature stage}.
Each single-electron source is operated at $f\sn{=}\qty{200}{MHz}$, which generates current of $ef\simeq\qty{32}{pA}$.
Current from single-electron sources is measured by a digital volt meter and transimpedance amplifier (Femto DDPCA-300) connected to the source terminal.
The other end, the drain terminal in the middle, is grounded at room temperature.

\begin{figure}[htbp]
\centering
\includegraphics[width=3.2in]{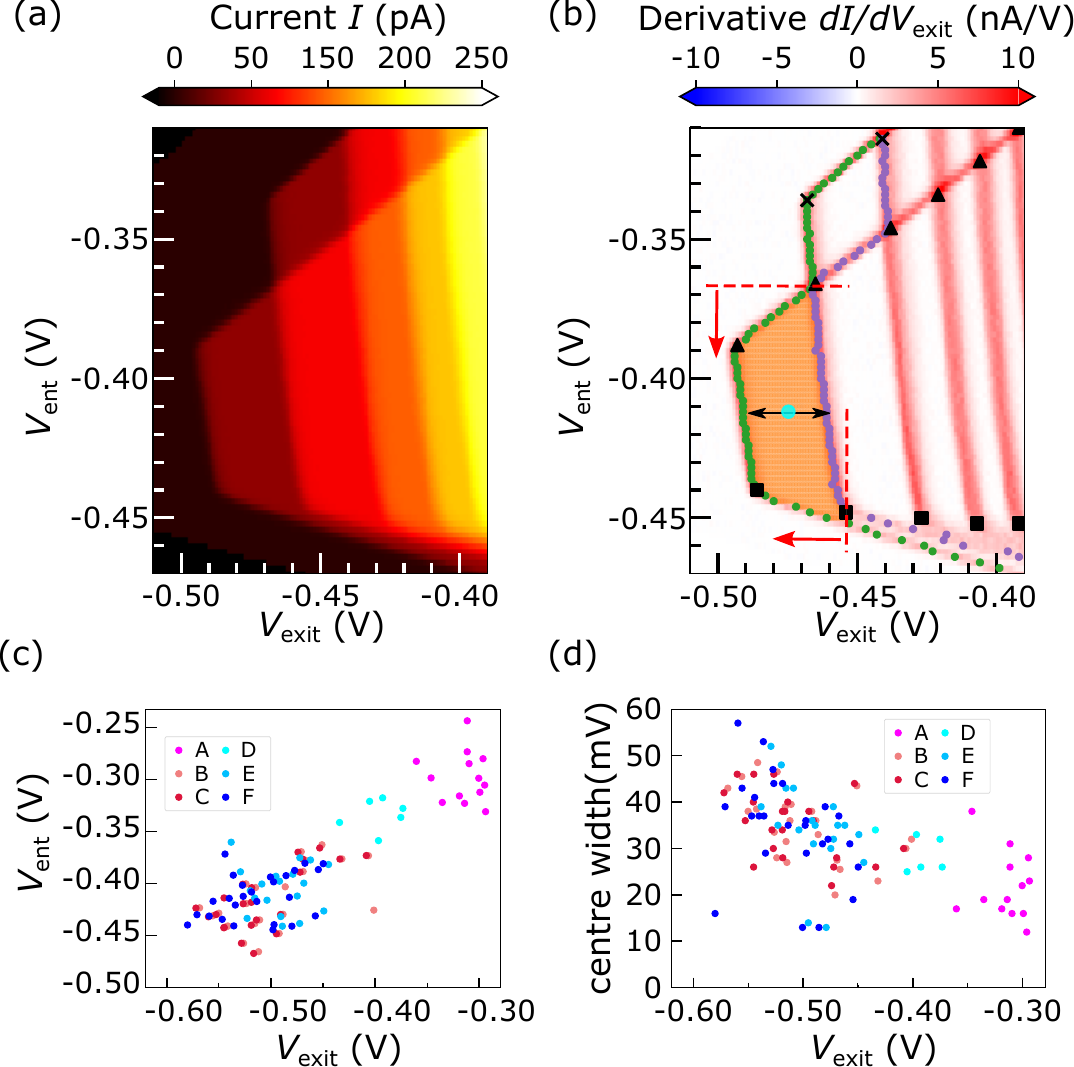}%
\caption{(a) Pumped current as functions of $V_{\textrm{ent}}$ and $V_{\textrm{exit}}$. (b) The derivative of pumped current with respect to $V_{\textrm{exit}}$ as functions of $V_{\textrm{ent}}$ and $V_{\textrm{exit}}$. The orange-\mk{filled} region indicates $1ef$ plateau. Black squares, triangles and crosses show plateau edges to detect the \mkk{$1ef$} plateau by our algorithm. \mkk{Red horizontal and vertical dashed lines and arrows indicate a condition that we used to assign the $1ef$ plateau.} A cyan bullet and black arrow are the centre of the $1ef$ plateau and its width, respectively. Two \mkk{chips of multiplexed} devices are measured to take statistics; Sample~1 at \qty{12.5}{T} (A-C) and Sample~2 at \qty{10.0}{T} (D-F). A and D are data for devices with emission direction \mkk{along} $[0\overline{11}]$. The rest are devices with emission direction \mkk{along} $[0\overline{1}1]$. C and F are data that a stronger microwave is applied than B (\qty{+0.5}{dB}) and E (\qty{+1.6}{dB}), respectively. (c) A distribution for \mkk{$1ef$} plateau centre in a space of $V_{\textrm{ent}}$ and $V_{\textrm{exit}}$. (d) A distribution for \mkk{$1ef$} plateau width as a function of the centre position in $V_{\textrm{exit}}$.\label{stat}}
\end{figure}
Figure~\ref{stat}(a, b) show a typical pumped current and its derivative with respect to $V_{\textrm{exit}}$ as functions of $V_{\textrm{ent}}$ and $V_{\textrm{exit}}$, respectively.
The orange region and cyan bullet in Fig.~\ref{stat}(b) show the $1ef$ plateau and its centre.
We define the centre position as the centre of mass \mk{for the} $1ef$ plateau, \mk{which is found by the following automated method}.
In order to process a large number of datasets produced by multiplexed pumps, we use an automated algorithm to extract the $1ef$ plateau region.
Our algorithm follows several steps, finding its boundary, capturing corners of the boundary and counting data points inside of the boundary.
We define $1ef$ plateau boundary using the current derivative with respect to $V_{\textrm{exit}}$.
Each point on the boundary is detected by finding the derivative maxima in the region where the current is in the range of $[0ef, 1ef]$ and $[1ef, 2ef]$ at each $V_{\textrm{ent}}$.
The boundaries in the current range of $[0ef, 1ef]$ and $[1ef, 2ef]$, transition lines, are indicated by green dots and purple dots in Fig.~\ref{stat}(b), respectively.
A naive definition that $1ef$ plateau is a region surrounded by the boundary would fail as the centre of $1ef$ plateau would be found outside of the plateau due to the narrow stripe of not-well-quantised $1ef$ plateau region extending towards the bottom right region in Fig.~\ref{stat}(b).
We define the area included in $1ef$ plateau to capture the centre of the plateau in terms of practical multi-operation.
The area is defined by corners of transition lines that are detected by evaluating orthogonal projections onto an appropriate line.
\mk{Black squares, triangles and crosses indicate the edges of the plateau found by our algorithm.}
\mkk{For the $1ef$ plateau, we include the points within the boundary that has the $V_{\textrm{ent}}$ value more negative than the top left corner (triangle) of the $1ef - 2ef$ transition line and the $V_{\textrm{exit}}$ value more negative than  the bottom left corner (square) of the same transition [see the red dashed lines and arrows in Fig.~\ref{stat}(b)].}

We analysed data that provide typical device characteristics shown in Fig.~\ref{stat}(a, b) to acquire statistics on $1ef$ plateau for two \mkk{chips of multiplexed} devices (Sample~1 and Sample~2).
Data points are categorised into group A-F depending on the device geometry and operation parameters.
Group A and D are devices emitting electrons towards $[0\overline{11}]$ in Sample~1 and Sample~2, respectively.
Group B and E are devices emitting electrons towards $[0\overline{1}1]$ in Sample~1 and Sample~2, respectively.
Group C and F are devices \mkk{emitting electrons as} in \mkk{Groups} B and E \mkk{but} operated by \qty{1.6}{dB} and \qty{0.5}{dB} stronger microwave, respectively. 
Figure~\ref{stat}(c) shows \mkk{the} distribution of $1ef$ plateau centre position in \mk{the} space of $V_{\textrm{ent}}$ and $V_{\textrm{exit}}$.
It is found that \mkk{operation with a stronger} microwave does not move the centre of the plateau dramatically.
\mkk{This can be seen by comparing the distribution of Groups B and C, and Groups E and F.}
\mkk{Operation with stronger microwave drives expands the $1ef$ plateau vertically, which leads to a slight negative shift of the plateau centre positions in the $V_{\textrm{ent}}$ and $V_{\textrm{exit}}$ space (\mk{by a small amount roughly the size of the symbol in most cases})}.
It is also found that the centre distributions strongly depend on electron emission direction in a device.
Single electron sources along $[0\overline{11}]$, namely \mkk{Groups} A and D, show a trend to have $1ef$ \mkk{plateau centre relatively} less negative voltages in \mk{the} space of $V_{\textrm{ent}}$ and $V_{\textrm{exit}}$ compared with devices along $[0\overline{1}1]$ (\mkk{Groups} B, C, E and F).
\mk{We also evaluate \mkk{the} $1ef$ plateau widths in $V_{\textrm{exit}}$ at the $V_{\textrm{ent}}$ going across the plateau centres [see Fig.~\ref{stat}(b)].
Figure~\ref{stat}(d) shows the distribution of $1ef$ plateau width as a function of the centre position in $V_{\textrm{exit}}$.
It is found that \mkk{the $1ef$ plateau width is distributed} in the range of [\qty{10}{mV}, \qty{60}{mV}] and the width is insensitive to microwave power.
The devices along $[0\overline{11}]$ tend to have shorter plateaus \mk{in the $V_{\textrm{exit}}$ direction}, which poses a disadvantage in our pump parallelisation scheme as finding common device operation parameters becomes less likely}.
For parallelisation, we chose to focus on devices in the $[0\overline{1}1]$ direction as their longer plateau is advantageous.

\mk{Because our single-electron sources share the \mkk{common} Entrance and Exit gates, our parallelisation scheme relies on chance to find single-electron sources that can be operated with common parameters for generating a quantised current. Also, these sources need to be in paths that are compatible with our multiplexer parallelisation scheme as described in Fig.~\ref{device}(a).}

We individually measured 64 single-electron sources in Sample 1 under the same \mk{microwave power and frequency to find a group} of devices to maximise current under the multiplexer limitation.
It \mk{was} found that \mk{the devices} 17, 19 and 49 shared their $1ef$ plateau region and \mk{the} device 51 generates no current under the condition.
The multiplexer in the state "X100X0" activates \mk{the} devices 17, 19, 49 and 51 simultaneously.
It is expected to find $3ef$ plateau consisting of $1ef$ plateaus for \mk{the devices} 17, 19 and 49, and no current from \mk{the} device 51.

\begin{figure}[htbp]
\centering
\includegraphics[width=3.2in]{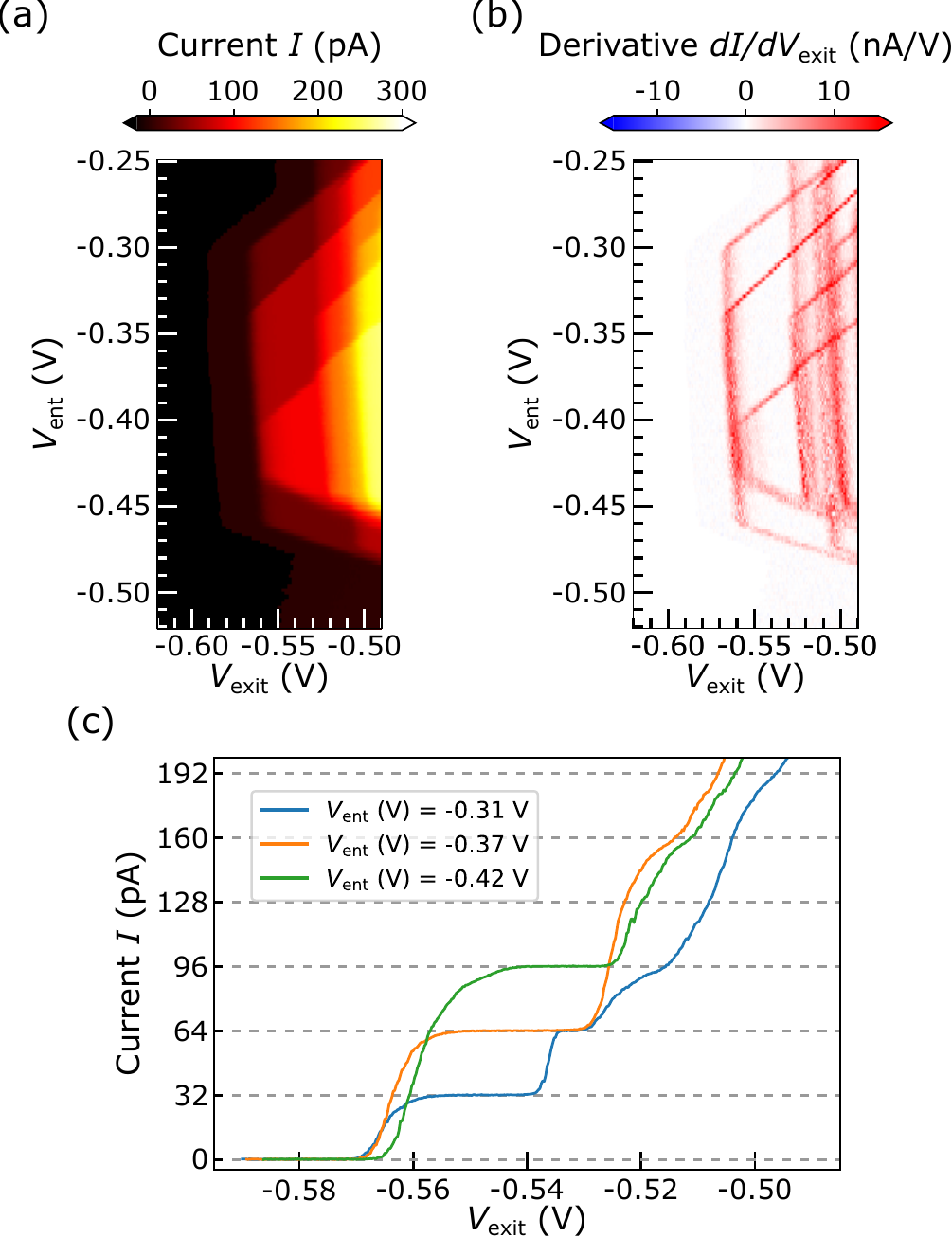}%
\caption{
        (a) Current as functions of $V_{\textrm{exit}}$ and $V_{\textrm{ent}}$ when the multiplexer accesses device 17, 19, 49 and 51 simultaneously.
        (b) Derivatives of generated current with respect to $V_{\textrm{exit}}$ as functions of $V_{\textrm{exit}}$ and $V_{\textrm{ent}}$ when the multiplexer accesses device 17, 19, 49 and 51 simultaneously.
        (c) Line profiles of the current at $V_{\textrm{ent}}=\qtylist{-0.31;-0.37;-0.42}{V}$. \label{parallel}}%
\end{figure}

\mkk{Figures~\ref{parallel}(a, b) show} the current from \mk{the} devices 17, 19, 49 and 51, operated in parallel, and its derivative with respect to $V_{\textrm{exit}}$ as functions of $V_{\textrm{ent}}$ and $V_{\textrm{exit}}$.
The region is observed that three $1ef$ plateaus overlap with each other.
Due to the variety in $1ef$ plateau size, there are several operation points where a different number of devices contributes to the current.
Figure~\ref{parallel}(c) provides line profiles of the current at $V_{\textrm{ent}}=\qty{-0.31}{V}$ (only \mk{the} device 17 contributes), \qty{-0.37}{V} (\mk{the devices} 17 and 49) and \qty{-0.42}{V} (\mk{the devices} 17, 19 and 49), respectively.
The horizontal grids appear every $1ef\simeq\qty{32}{pA}$.
It is found that the current increases by a factor of the number of devices contributing current generation.
\mkk{The} $3ef$ plateau is observed at $V_{\textrm{ent}}=\qty{-0.42}{V}$.
\mk{This indicates that, even though all single-electron sources are operated by common gates, activating multiple channels does not disrupt the \mkk{operation of} single-electron sources, and that the current \mkk{simply adds} up.}

We have demonstrated operation of multiplexed single-electron pumps.
We performed statistics for the device operation conditions for $1ef$ current quantisation.
It is found that the distribution of $1ef$ plateau \mkk{centres} strongly depends on electron emission direction and that single-electron sources aligned with $[0\overline{1}1]$ showed a trend to have longer plateaus.
A new device design \mkk{with all single-electron sources along} $[0\overline{1}1]$ would provide a chance for more devices to share $1ef$ plateaus and generate a larger amount of electric current.
We then demonstrated parallelisation of three single-electron pumps using the multiplexer circuit and that the generated current increases by a factor of three.
Our parallelisation scheme relies on \mk{a} coincidental agreement in device operation conditions for a certain combination that the multiplexer circuit allows to address simultaneously.
Due to the selection limitation, our method is not an efficient way to parallelise a larger number of single-electron sources.
The limitation, however, could be cleared by a \mk{charge-locking} technique that provides the capability to access an arbitrary group of devices~\cite{puddy2015}, which increases generated current by a factor of the number of devices sharing $1ef$ plateau.
These improvements would realise a standard for the ampere defined by the fixed numerical value of elementary charge and the \mkk{low uncertainty of the operating} frequency.

\section{Acknowledgements}
This work was supported by the UK government's Department for Science, Innovation and Technology\sn{, and a National Institute of Standards and Technology cooperative agreement 70NANB16H168}.
\sn{Certain commercial equipment, instruments, or materials are identified in this article to specify the experimental procedure adequately.}
\sn{Such identification is not intended to imply recommendation or endorsement by NIST, nor is it intended to imply that the materials or equipment identified are necessarily the best available for the purpose.}
\bibstyle{unsrt}


\begin{thebibliography}{16}%
\makeatletter
\providecommand \@ifxundefined [1]{%
 \@ifx{#1\undefined}
}%
\providecommand \@ifnum [1]{%
 \ifnum #1\expandafter \@firstoftwo
 \else \expandafter \@secondoftwo
 \fi
}%
\providecommand \@ifx [1]{%
 \ifx #1\expandafter \@firstoftwo
 \else \expandafter \@secondoftwo
 \fi
}%
\providecommand \natexlab [1]{#1}%
\providecommand \enquote  [1]{``#1''}%
\providecommand \bibnamefont  [1]{#1}%
\providecommand \bibfnamefont [1]{#1}%
\providecommand \citenamefont [1]{#1}%
\providecommand \href@noop [0]{\@secondoftwo}%
\providecommand \href [0]{\begingroup \@sanitize@url \@href}%
\providecommand \@href[1]{\@@startlink{#1}\@@href}%
\providecommand \@@href[1]{\endgroup#1\@@endlink}%
\providecommand \@sanitize@url [0]{\catcode `\\12\catcode `\$12\catcode
  `\&12\catcode `\#12\catcode `\^12\catcode `\_12\catcode `\%12\relax}%
\providecommand \@@startlink[1]{}%
\providecommand \@@endlink[0]{}%
\providecommand \url  [0]{\begingroup\@sanitize@url \@url }%
\providecommand \@url [1]{\endgroup\@href {#1}{\urlprefix }}%
\providecommand \urlprefix  [0]{URL }%
\providecommand \Eprint [0]{\href }%
\providecommand \doibase [0]{http://dx.doi.org/}%
\providecommand \selectlanguage [0]{\@gobble}%
\providecommand \bibinfo  [0]{\@secondoftwo}%
\providecommand \bibfield  [0]{\@secondoftwo}%
\providecommand \translation [1]{[#1]}%
\providecommand \BibitemOpen [0]{}%
\providecommand \bibitemStop [0]{}%
\providecommand \bibitemNoStop [0]{.\EOS\space}%
\providecommand \EOS [0]{\spacefactor3000\relax}%
\providecommand \BibitemShut  [1]{\csname bibitem#1\endcsname}%
\let\auto@bib@innerbib\@empty
\bibitem [{\citenamefont {des Poids~et Measures}(2019)}]{bipm2019}%
  \BibitemOpen
  \bibfield  {author} {\bibinfo {author} {\bibfnamefont {B.~I.}\ \bibnamefont
  {des Poids~et Measures}},\ }\href@noop {} {\enquote {\bibinfo {title} {The
  international system of units 9th edition},}\ } (\bibinfo {year}
  {2019})\BibitemShut {NoStop}%
\bibitem [{\citenamefont {Brun-Picard}\ \emph {et~al.}(2016)\citenamefont
  {Brun-Picard}, \citenamefont {Djordjevic}, \citenamefont {Leprat},
  \citenamefont {Schopfer},\ and\ \citenamefont {Poirier}}]{Brun2016}%
  \BibitemOpen
  \bibfield  {author} {\bibinfo {author} {\bibfnamefont {J.}~\bibnamefont
  {Brun-Picard}}, \bibinfo {author} {\bibfnamefont {S.}~\bibnamefont
  {Djordjevic}}, \bibinfo {author} {\bibfnamefont {D.}~\bibnamefont {Leprat}},
  \bibinfo {author} {\bibfnamefont {F.}~\bibnamefont {Schopfer}}, \ and\
  \bibinfo {author} {\bibfnamefont {W.}~\bibnamefont {Poirier}},\ }\bibfield
  {title} {\enquote {\bibinfo {title} {Practical quantum realization of the
  ampere from the elementary charge},}\ }\href {\doibase
  10.1103/PhysRevX.6.041051} {\bibfield  {journal} {\bibinfo  {journal} {Phys.
  Rev. X}\ }\textbf {\bibinfo {volume} {6}},\ \bibinfo {pages} {041051}
  (\bibinfo {year} {2016})}\BibitemShut {NoStop}%
\bibitem [{\citenamefont {Pekola}\ \emph {et~al.}(2013)\citenamefont {Pekola},
  \citenamefont {Saira}, \citenamefont {Maisi}, \citenamefont {Kemppinen},
  \citenamefont {M\"ott\"onen}, \citenamefont {Pashkin},\ and\ \citenamefont
  {Averin}}]{jukka2013}%
  \BibitemOpen
  \bibfield  {author} {\bibinfo {author} {\bibfnamefont {J.~P.}\ \bibnamefont
  {Pekola}}, \bibinfo {author} {\bibfnamefont {O.-P.}\ \bibnamefont {Saira}},
  \bibinfo {author} {\bibfnamefont {V.~F.}\ \bibnamefont {Maisi}}, \bibinfo
  {author} {\bibfnamefont {A.}~\bibnamefont {Kemppinen}}, \bibinfo {author}
  {\bibfnamefont {M.}~\bibnamefont {M\"ott\"onen}}, \bibinfo {author}
  {\bibfnamefont {Y.~A.}\ \bibnamefont {Pashkin}}, \ and\ \bibinfo {author}
  {\bibfnamefont {D.~V.}\ \bibnamefont {Averin}},\ }\bibfield  {title}
  {\enquote {\bibinfo {title} {Single-electron current sources: Toward a
  refined definition of the ampere},}\ }\href {\doibase
  10.1103/RevModPhys.85.1421} {\bibfield  {journal} {\bibinfo  {journal} {Rev.
  Mod. Phys.}\ }\textbf {\bibinfo {volume} {85}},\ \bibinfo {pages}
  {1421--1472} (\bibinfo {year} {2013})}\BibitemShut {NoStop}%
\bibitem [{\citenamefont {Kaneko}, \citenamefont {Nakamura},\ and\
  \citenamefont {Okazaki}(2016)}]{Kaneko_2016}%
  \BibitemOpen
  \bibfield  {author} {\bibinfo {author} {\bibfnamefont {N.-H.}\ \bibnamefont
  {Kaneko}}, \bibinfo {author} {\bibfnamefont {S.}~\bibnamefont {Nakamura}}, \
  and\ \bibinfo {author} {\bibfnamefont {Y.}~\bibnamefont {Okazaki}},\
  }\bibfield  {title} {\enquote {\bibinfo {title} {A review of the quantum
  current standard},}\ }\href {\doibase 10.1088/0957-0233/27/3/032001}
  {\bibfield  {journal} {\bibinfo  {journal} {Meas. Sci. Technol.}\ }\textbf
  {\bibinfo {volume} {27}},\ \bibinfo {pages} {032001} (\bibinfo {year}
  {2016})}\BibitemShut {NoStop}%
\bibitem [{\citenamefont {Kataoka}\ \emph {et~al.}(2011)\citenamefont
  {Kataoka}, \citenamefont {Fletcher}, \citenamefont {See}, \citenamefont
  {Giblin}, \citenamefont {Janssen}, \citenamefont {Griffiths}, \citenamefont
  {Jones}, \citenamefont {Farrer},\ and\ \citenamefont
  {Ritchie}}]{Kataoka_2011}%
  \BibitemOpen
  \bibfield  {author} {\bibinfo {author} {\bibfnamefont {M.}~\bibnamefont
  {Kataoka}}, \bibinfo {author} {\bibfnamefont {J.~D.}\ \bibnamefont
  {Fletcher}}, \bibinfo {author} {\bibfnamefont {P.}~\bibnamefont {See}},
  \bibinfo {author} {\bibfnamefont {S.~P.}\ \bibnamefont {Giblin}}, \bibinfo
  {author} {\bibfnamefont {T.~J. B.~M.}\ \bibnamefont {Janssen}}, \bibinfo
  {author} {\bibfnamefont {J.~P.}\ \bibnamefont {Griffiths}}, \bibinfo {author}
  {\bibfnamefont {G.~A.~C.}\ \bibnamefont {Jones}}, \bibinfo {author}
  {\bibfnamefont {I.}~\bibnamefont {Farrer}}, \ and\ \bibinfo {author}
  {\bibfnamefont {D.~A.}\ \bibnamefont {Ritchie}},\ }\bibfield  {title}
  {\enquote {\bibinfo {title} {Tunable nonadiabatic excitation in a
  single-electron quantum dot},}\ }\href {\doibase
  10.1103/PhysRevLett.106.126801} {\bibfield  {journal} {\bibinfo  {journal}
  {Phys. Rev. Lett.}\ }\textbf {\bibinfo {volume} {106}},\ \bibinfo {pages}
  {126801} (\bibinfo {year} {2011})}\BibitemShut {NoStop}%
\bibitem [{\citenamefont {Yamahata}\ \emph {et~al.}(2017)\citenamefont
  {Yamahata}, \citenamefont {Giblin}, \citenamefont {Kataoka}, \citenamefont
  {Karasawa},\ and\ \citenamefont {Fujiwara}}]{Yamahata2017-nw}%
  \BibitemOpen
  \bibfield  {author} {\bibinfo {author} {\bibfnamefont {G.}~\bibnamefont
  {Yamahata}}, \bibinfo {author} {\bibfnamefont {S.~P.}\ \bibnamefont
  {Giblin}}, \bibinfo {author} {\bibfnamefont {M.}~\bibnamefont {Kataoka}},
  \bibinfo {author} {\bibfnamefont {T.}~\bibnamefont {Karasawa}}, \ and\
  \bibinfo {author} {\bibfnamefont {A.}~\bibnamefont {Fujiwara}},\ }\bibfield
  {title} {\enquote {\bibinfo {title} {High-accuracy current generation in the
  nanoampere regime from a silicon single-trap electron pump},}\ }\href@noop {}
  {\bibfield  {journal} {\bibinfo  {journal} {Sci. Rep.}\ }\textbf {\bibinfo
  {volume} {7}},\ \bibinfo {pages} {45137} (\bibinfo {year}
  {2017})}\BibitemShut {NoStop}%
\bibitem [{\citenamefont {Giblin}\ \emph {et~al.}(2023)\citenamefont {Giblin},
  \citenamefont {Yamahata}, \citenamefont {Fujiwara},\ and\ \citenamefont
  {Kataoka}}]{Giblin_2023}%
  \BibitemOpen
  \bibfield  {author} {\bibinfo {author} {\bibfnamefont {S.~P.}\ \bibnamefont
  {Giblin}}, \bibinfo {author} {\bibfnamefont {G.}~\bibnamefont {Yamahata}},
  \bibinfo {author} {\bibfnamefont {A.}~\bibnamefont {Fujiwara}}, \ and\
  \bibinfo {author} {\bibfnamefont {M.}~\bibnamefont {Kataoka}},\ }\bibfield
  {title} {\enquote {\bibinfo {title} {Precision measurement of an electron
  pump at 2~ghz; the frontier of small dc current metrology},}\ }\href
  {\doibase 10.1088/1681-7575/ace054} {\bibfield  {journal} {\bibinfo
  {journal} {Metrologia}\ }\textbf {\bibinfo {volume} {60}},\ \bibinfo {pages}
  {055001} (\bibinfo {year} {2023})}\BibitemShut {NoStop}%
\bibitem [{\citenamefont {Maisi}\ \emph {et~al.}(2009)\citenamefont {Maisi},
  \citenamefont {Pashkin}, \citenamefont {Kafanov}, \citenamefont {Tsai},\ and\
  \citenamefont {Pekola}}]{Maisi_2009}%
  \BibitemOpen
  \bibfield  {author} {\bibinfo {author} {\bibfnamefont {V.~F.}\ \bibnamefont
  {Maisi}}, \bibinfo {author} {\bibfnamefont {Y.~A.}\ \bibnamefont {Pashkin}},
  \bibinfo {author} {\bibfnamefont {S.}~\bibnamefont {Kafanov}}, \bibinfo
  {author} {\bibfnamefont {J.-S.}\ \bibnamefont {Tsai}}, \ and\ \bibinfo
  {author} {\bibfnamefont {J.~P.}\ \bibnamefont {Pekola}},\ }\bibfield  {title}
  {\enquote {\bibinfo {title} {Parallel pumping of electrons},}\ }\href
  {\doibase 10.1088/1367-2630/11/11/113057} {\bibfield  {journal} {\bibinfo
  {journal} {New J. Phys.}\ }\textbf {\bibinfo {volume} {11}},\ \bibinfo
  {pages} {113057} (\bibinfo {year} {2009})}\BibitemShut {NoStop}%
\bibitem [{\citenamefont {Nakamura}\ \emph {et~al.}(2015)\citenamefont
  {Nakamura}, \citenamefont {Pashkin}, \citenamefont {Tsai},\ and\
  \citenamefont {Kaneko}}]{shuji2015}%
  \BibitemOpen
  \bibfield  {author} {\bibinfo {author} {\bibfnamefont {S.}~\bibnamefont
  {Nakamura}}, \bibinfo {author} {\bibfnamefont {Y.~A.}\ \bibnamefont
  {Pashkin}}, \bibinfo {author} {\bibfnamefont {J.-S.}\ \bibnamefont {Tsai}}, \
  and\ \bibinfo {author} {\bibfnamefont {N.-h.}\ \bibnamefont {Kaneko}},\
  }\bibfield  {title} {\enquote {\bibinfo {title} {Single-electron pumping by
  parallel sinis turnstiles for quantum current standard},}\ }\href {\doibase
  10.1109/TIM.2015.2418452} {\bibfield  {journal} {\bibinfo  {journal} {IEEE
  Trans. Instrum. Meas.}\ }\textbf {\bibinfo {volume} {64}},\ \bibinfo {pages}
  {1696--1701} (\bibinfo {year} {2015})}\BibitemShut {NoStop}%
\bibitem [{\citenamefont {Nakamura}\ \emph {et~al.}(2023)\citenamefont
  {Nakamura}, \citenamefont {Matsumaru}, \citenamefont {Yamahata},
  \citenamefont {Oe}, \citenamefont {Chae}, \citenamefont {Okazaki},
  \citenamefont {Takada}, \citenamefont {Maruyama}, \citenamefont {Fujiwara},\
  and\ \citenamefont {Kaneko}}]{Nakamura2023-tx}%
  \BibitemOpen
  \bibfield  {author} {\bibinfo {author} {\bibfnamefont {S.}~\bibnamefont
  {Nakamura}}, \bibinfo {author} {\bibfnamefont {D.}~\bibnamefont {Matsumaru}},
  \bibinfo {author} {\bibfnamefont {G.}~\bibnamefont {Yamahata}}, \bibinfo
  {author} {\bibfnamefont {T.}~\bibnamefont {Oe}}, \bibinfo {author}
  {\bibfnamefont {D.-H.}\ \bibnamefont {Chae}}, \bibinfo {author}
  {\bibfnamefont {Y.}~\bibnamefont {Okazaki}}, \bibinfo {author} {\bibfnamefont
  {S.}~\bibnamefont {Takada}}, \bibinfo {author} {\bibfnamefont
  {M.}~\bibnamefont {Maruyama}}, \bibinfo {author} {\bibfnamefont
  {A.}~\bibnamefont {Fujiwara}}, \ and\ \bibinfo {author} {\bibfnamefont
  {N.-H.}\ \bibnamefont {Kaneko}},\ }\bibfield  {title} {\enquote {\bibinfo
  {title} {Universality and multiplication of {Gigahertz-Operated} silicon
  pumps with parts per {Million-Level} uncertainty},}\ }\href@noop {}
  {\bibfield  {journal} {\bibinfo  {journal} {Nano Lett.}\ }\textbf {\bibinfo
  {volume} {24}},\ \bibinfo {pages} {9--15} (\bibinfo {year}
  {2023})}\BibitemShut {NoStop}%
\bibitem [{\citenamefont {Mirovsky}\ \emph {et~al.}(2010)\citenamefont
  {Mirovsky}, \citenamefont {Kaestner}, \citenamefont {Leicht}, \citenamefont
  {Welker}, \citenamefont {Weimann}, \citenamefont {Pierz},\ and\ \citenamefont
  {Schumacher}}]{Mirovsky_2010}%
  \BibitemOpen
  \bibfield  {author} {\bibinfo {author} {\bibfnamefont {P.}~\bibnamefont
  {Mirovsky}}, \bibinfo {author} {\bibfnamefont {B.}~\bibnamefont {Kaestner}},
  \bibinfo {author} {\bibfnamefont {C.}~\bibnamefont {Leicht}}, \bibinfo
  {author} {\bibfnamefont {A.~C.}\ \bibnamefont {Welker}}, \bibinfo {author}
  {\bibfnamefont {T.}~\bibnamefont {Weimann}}, \bibinfo {author} {\bibfnamefont
  {K.}~\bibnamefont {Pierz}}, \ and\ \bibinfo {author} {\bibfnamefont {H.~W.}\
  \bibnamefont {Schumacher}},\ }\bibfield  {title} {\enquote {\bibinfo {title}
  {{Synchronized single electron emission from dynamical quantum dots}},}\
  }\href {\doibase 10.1063/1.3527940} {\bibfield  {journal} {\bibinfo
  {journal} {Applied Physics Letters}\ }\textbf {\bibinfo {volume} {97}},\
  \bibinfo {pages} {252104} (\bibinfo {year} {2010})},\ \Eprint
  {http://arxiv.org/abs/https://pubs.aip.org/aip/apl/article-pdf/doi/10.1063/1.3527940/13620115/252104\_1\_online.pdf}
  {https://pubs.aip.org/aip/apl/article-pdf/doi/10.1063/1.3527940/13620115/252104\_1\_online.pdf}
  \BibitemShut {NoStop}%
\bibitem [{\citenamefont {Kim}\ \emph {et~al.}(2022)\citenamefont {Kim},
  \citenamefont {Yu}, \citenamefont {Park}, \citenamefont {Song}, \citenamefont
  {Kim},\ and\ \citenamefont {Bae}}]{kim2022}%
  \BibitemOpen
  \bibfield  {author} {\bibinfo {author} {\bibfnamefont {B.-K.}\ \bibnamefont
  {Kim}}, \bibinfo {author} {\bibfnamefont {B.-S.}\ \bibnamefont {Yu}},
  \bibinfo {author} {\bibfnamefont {S.-I.}\ \bibnamefont {Park}}, \bibinfo
  {author} {\bibfnamefont {J.}~\bibnamefont {Song}}, \bibinfo {author}
  {\bibfnamefont {N.}~\bibnamefont {Kim}}, \ and\ \bibinfo {author}
  {\bibfnamefont {M.-H.}\ \bibnamefont {Bae}},\ }\bibfield  {title} {\enquote
  {\bibinfo {title} {{Tuning current plateau regions in parallelized
  single-electron pumps}},}\ }\href {\doibase 10.1063/5.0117055} {\bibfield
  {journal} {\bibinfo  {journal} {AIP Adv.}\ }\textbf {\bibinfo {volume}
  {12}},\ \bibinfo {pages} {105118} (\bibinfo {year} {2022})}\BibitemShut
  {NoStop}%
\bibitem [{\citenamefont {Al-Taie}\ \emph {et~al.}(2013)\citenamefont
  {Al-Taie}, \citenamefont {Smith}, \citenamefont {Xu}, \citenamefont {See},
  \citenamefont {Griffiths}, \citenamefont {Beere}, \citenamefont {Jones},
  \citenamefont {Ritchie}, \citenamefont {Kelly},\ and\ \citenamefont
  {Smith}}]{al2013}%
  \BibitemOpen
  \bibfield  {author} {\bibinfo {author} {\bibfnamefont {H.}~\bibnamefont
  {Al-Taie}}, \bibinfo {author} {\bibfnamefont {L.~W.}\ \bibnamefont {Smith}},
  \bibinfo {author} {\bibfnamefont {B.}~\bibnamefont {Xu}}, \bibinfo {author}
  {\bibfnamefont {P.}~\bibnamefont {See}}, \bibinfo {author} {\bibfnamefont
  {J.~P.}\ \bibnamefont {Griffiths}}, \bibinfo {author} {\bibfnamefont {H.~E.}\
  \bibnamefont {Beere}}, \bibinfo {author} {\bibfnamefont {G.~A.~C.}\
  \bibnamefont {Jones}}, \bibinfo {author} {\bibfnamefont {D.~A.}\ \bibnamefont
  {Ritchie}}, \bibinfo {author} {\bibfnamefont {M.~J.}\ \bibnamefont {Kelly}},
  \ and\ \bibinfo {author} {\bibfnamefont {C.~G.}\ \bibnamefont {Smith}},\
  }\bibfield  {title} {\enquote {\bibinfo {title} {{Cryogenic on-chip
  multiplexer for the study of quantum transport in 256 split-gate devices}},}\
  }\href {\doibase 10.1063/1.4811376} {\bibfield  {journal} {\bibinfo
  {journal} {Appl. Phys. Lett.}\ }\textbf {\bibinfo {volume} {102}},\ \bibinfo
  {pages} {243102} (\bibinfo {year} {2013})}\BibitemShut {NoStop}%
\bibitem [{\citenamefont {Blumenthal}\ \emph {et~al.}(2007)\citenamefont
  {Blumenthal}, \citenamefont {Kaestner}, \citenamefont {Li}, \citenamefont
  {Giblin}, \citenamefont {Janssen}, \citenamefont {Pepper}, \citenamefont
  {Anderson}, \citenamefont {Jones},\ and\ \citenamefont
  {Ritchie}}]{Blumenthal2007-ki}%
  \BibitemOpen
  \bibfield  {author} {\bibinfo {author} {\bibfnamefont {M.~D.}\ \bibnamefont
  {Blumenthal}}, \bibinfo {author} {\bibfnamefont {B.}~\bibnamefont
  {Kaestner}}, \bibinfo {author} {\bibfnamefont {L.}~\bibnamefont {Li}},
  \bibinfo {author} {\bibfnamefont {S.}~\bibnamefont {Giblin}}, \bibinfo
  {author} {\bibfnamefont {T.~J. B.~M.}\ \bibnamefont {Janssen}}, \bibinfo
  {author} {\bibfnamefont {M.}~\bibnamefont {Pepper}}, \bibinfo {author}
  {\bibfnamefont {D.}~\bibnamefont {Anderson}}, \bibinfo {author}
  {\bibfnamefont {G.}~\bibnamefont {Jones}}, \ and\ \bibinfo {author}
  {\bibfnamefont {D.~A.}\ \bibnamefont {Ritchie}},\ }\bibfield  {title}
  {\enquote {\bibinfo {title} {Gigahertz quantized charge pumping},}\
  }\href@noop {} {\bibfield  {journal} {\bibinfo  {journal} {Nat. Phys.}\
  }\textbf {\bibinfo {volume} {3}},\ \bibinfo {pages} {343--347} (\bibinfo
  {year} {2007})}\BibitemShut {NoStop}%
\bibitem [{\citenamefont {Giblin}\ \emph {et~al.}(2012)\citenamefont {Giblin},
  \citenamefont {Kataoka}, \citenamefont {Fletcher}, \citenamefont {See},
  \citenamefont {Janssen}, \citenamefont {Griffiths}, \citenamefont {Jones},
  \citenamefont {Farrer},\ and\ \citenamefont {Ritchie}}]{Giblin2012-wf}%
  \BibitemOpen
  \bibfield  {author} {\bibinfo {author} {\bibfnamefont {S.~P.}\ \bibnamefont
  {Giblin}}, \bibinfo {author} {\bibfnamefont {M.}~\bibnamefont {Kataoka}},
  \bibinfo {author} {\bibfnamefont {J.~D.}\ \bibnamefont {Fletcher}}, \bibinfo
  {author} {\bibfnamefont {P.}~\bibnamefont {See}}, \bibinfo {author}
  {\bibfnamefont {T.~J. B.~M.}\ \bibnamefont {Janssen}}, \bibinfo {author}
  {\bibfnamefont {J.~P.}\ \bibnamefont {Griffiths}}, \bibinfo {author}
  {\bibfnamefont {G.~A.~C.}\ \bibnamefont {Jones}}, \bibinfo {author}
  {\bibfnamefont {I.}~\bibnamefont {Farrer}}, \ and\ \bibinfo {author}
  {\bibfnamefont {D.~A.}\ \bibnamefont {Ritchie}},\ }\bibfield  {title}
  {\enquote {\bibinfo {title} {Towards a quantum representation of the ampere
  using single electron pumps},}\ }\href@noop {} {\bibfield  {journal}
  {\bibinfo  {journal} {Nat. Commun.}\ }\textbf {\bibinfo {volume} {3}},\
  \bibinfo {pages} {930} (\bibinfo {year} {2012})}\BibitemShut {NoStop}%
\bibitem [{\citenamefont {Puddy}\ \emph {et~al.}(2015)\citenamefont {Puddy},
  \citenamefont {Smith}, \citenamefont {Al-Taie}, \citenamefont {Chong},
  \citenamefont {Farrer}, \citenamefont {Griffiths}, \citenamefont {Ritchie},
  \citenamefont {Kelly}, \citenamefont {Pepper},\ and\ \citenamefont
  {Smith}}]{puddy2015}%
  \BibitemOpen
  \bibfield  {author} {\bibinfo {author} {\bibfnamefont {R.~K.}\ \bibnamefont
  {Puddy}}, \bibinfo {author} {\bibfnamefont {L.~W.}\ \bibnamefont {Smith}},
  \bibinfo {author} {\bibfnamefont {H.}~\bibnamefont {Al-Taie}}, \bibinfo
  {author} {\bibfnamefont {C.~H.}\ \bibnamefont {Chong}}, \bibinfo {author}
  {\bibfnamefont {I.}~\bibnamefont {Farrer}}, \bibinfo {author} {\bibfnamefont
  {J.~P.}\ \bibnamefont {Griffiths}}, \bibinfo {author} {\bibfnamefont {D.~A.}\
  \bibnamefont {Ritchie}}, \bibinfo {author} {\bibfnamefont {M.~J.}\
  \bibnamefont {Kelly}}, \bibinfo {author} {\bibfnamefont {M.}~\bibnamefont
  {Pepper}}, \ and\ \bibinfo {author} {\bibfnamefont {C.~G.}\ \bibnamefont
  {Smith}},\ }\bibfield  {title} {\enquote {\bibinfo {title} {{Multiplexed
  charge-locking device for large arrays of quantum devices}},}\ }\href
  {\doibase 10.1063/1.4932012} {\bibfield  {journal} {\bibinfo  {journal}
  {Appl. Phys. Lett.}\ }\textbf {\bibinfo {volume} {107}},\ \bibinfo {pages}
  {143501} (\bibinfo {year} {2015})}\BibitemShut {NoStop}%
\end{thebibliography}
\end{document}